# A short review and discussion about the limiting factors, which can hold back wind and photovoltaic power plants from its presently exponential growth

Manfred G. Kratzenberg[1], Hans Helmut Zürn[2], Ricardo Rüther[1]

1 Fotovoltaica UFSC – Solar Energy Research Laboratory, Federal University of Santa Catarina, Florianópolis 88056-000, SC Brazil

2 Power System laboratory UFSC, Federal University of Santa Catarina, Campus Universitário Campus Universitário João David Ferreira Lima – Trindade 88040-900, Florianópolis, SC, Brasil.

The present nearly exponential growth of the cumulative installation power of wind and photovoltaic power plants is very promising in the context of a rapid reduction of the emission of $CO_2$, which can lead to a swift mitigation climate change if this growth behavior is maintained in the future. In this review we identify a set of ten limiting factors that can restrain, or halt back, an exponential growth of these variable renewable power plants (VREs) in the future. If the exponential growth slows down to a linear growth, such a case would extent considerably the time to reduce the $CO_2$ emissions and the related mitigation of climate change. A scenario that would result to a much higher risk of a future continuation of the economic grow and thrive of the global economy and human society. We argue that if these limiting factors are adequately addressed to an exponential expansion VRE power plants is feasible and can result in the nearly complete avoidance of $CO_2$ emissions as related to energy generation and consumption in 2030.

## 1. Introduction

As shown in (Kratzenberg, 2021) it is possible to reduce the emission of $CO_2$ as related to the electrical and primary energy generation to mitigate climate change much earlier as previously assumed or scheduled (Huang and Erb), (Kusch-Brandt, 2019) in 2050, if the present exponential growth of the cumulative installed power of wind and photovoltaic power plants is maintained in the near future. The resources of these power plants are much higher than the demanded electrical energy, and even primary energy for many decades in the future. Therefore, an exponential expansion of the generation capacity of these power plans is accomplishable, which is not the case of other renewable power plants, e.g. hydropower plants. Plants for the exploration of wave energy do also have this potential, but their commercial growth is still at an initial stage and, therefore, its exponential growth would not result in any significant contribution in the short-term time scale. However, an non-restrained further exponential expansion of wind and photovoltaic power plants would result in the complete avoidance of $CO_2$ emissions as related to the electrical and primary energy generation in 2030 as shown in (Kratzenberg, 2021). Such a configuration would result in an approximately 73 % reduction of the emissions of $CO_2$ in 2030, estimated on the relationship that presently 73.2 % of the world's $CO_2$ emissions are related to the generation of electrical and primary energy (Fawzy et al., 2020). Apart from (i) the strong reduction of $CO_2$ emissions, and (ii) a eventual swift mitigation of climate change, such an expansion also results in (iii) the beneficial generation of electrical and primary energy at much lower costs, and (iv) a large increase of employment, being all beneficial and important for a future sustainable development and thrive of the human society. However, there are several factors, which can results in considerable time lags of such an exponential expansion of the VRE generation, which we discuss in the following sections.

## 2.1 Manufacturing of VRE plants

We consider that there are no material or manufacturing constrictions for the exponential growth of photovoltaic power plants, as the most used silicon semiconductor is one of the most available materials on the earth's surface. Limits in the manufacturing ability of silicon photovoltaic modules were found solely in the past, when temporarily scarcity of solar cell grade silicon wafers were experienced. As the first production lines of PV modules were based on rejected wavers from the microchip manufacturing, its supply limited temporarily the expansion of this technology. The problem was solved by the delivery of a specific solar cell grade silicon, from which the polycrystalline cells were then manufactured on a large scale. The silicon waver scarcity led to a slight reduction of PV module manufacturing up to the year 2010. This reduced grow was, however, compensated by a stronger growth in the following years (Figure 1b in (Kratzenberg, 2021)). The rare earth metal (REM) neodymium, as used for permanent magnets in generators of aero generators, can affect the price of these devices. However, the generators of aero generators can also be manufactured without these magnets (Baldi et al., 2014). As a result, many companies are transitioning from providing electric generators in their wind turbines not based on neodymium. In these generators, necessary excitation of the magnetic field is supplied by electric currents, in combination with power electronics controlled by software (Alvarez, 2017). E.g., (i) the switched-reluctance generators (SRG), (Wu et al., 2018), (ii) the double feed induction generator (DFIG), (iii) the high-speed squirrel-cage (SCIG) induction generator, (iv) the electrically exited synchronous motor (EESG) (Pavel et al., 2017), but also the multiphase induction motor, as used in tesla EV (InsideEVs, 2021), do not use neodymium or windings in the core of the generator, but can present moderate performance losses in some functions (Pavel et al., 2017). E.g. the multiphase induction machine presents a ~ 4% efficiency loss when operated in EVs if compared to synchronous machine (InsideEVs, 2021). In the case of exponential growth of wind turbine generators, manufactures may prefer neodymium-free aero generators, depending on the neodymium's market price. The estimated 4% lower energy generation in turbines without permanent magnets, is not significant enough to avoid the penetration of wind turbines on an exponential scale because of the much higher general price decline.

### 2.2 Life expectancy and long-term maintenance cost

We consider the life expectancy of PV or wind power plants as no limiting factor for the exponential growth of the VRE generation. Life expectancy, or useful life, is for a PV module determined by the period at which it loses 20 % of its specified power under normalized solar irradiation and outdoor conditions. Because of the long term manufacturing experience with silicon cells and modules the degradation of its power conversion efficiency occurs at a slow pace. As a result of this low degradation tendency, photovoltaic modules' life expectancy is very high, from 25 to 30 years (Tantau and Staiger, 2017). However, in a PV system, the power converter presents another limiting factor with its life expectancy of approximately 10 to 15 years (Woodhouse et al., 2016). PV inverters are very rapidly replaceable, ideally by plug and play procedures. We recommend that PV projects should already consider a converter replacement and include it in the long-term corrective maintenance cost. The lifetime assessment of photovoltaic modules and systems is essential in the present context because low lifetimes, would result to large substitutions of PV panels to attain the fully renewable energy supply in 2032. However, because of the long PV-module lifetime of 25 to 30 years, substitution is only necessary for a tiny part of the PV-systems installed before 2002 or 2007. Compared to the peak power installed in 2002, the estimated peak power in 2032 is 12246 times higher. Therefore, the systems installed before 2002 do not have a significant weight in comparison.

### 2.3 Available resources

Enough high-potential installation locations for VRE systems are available, which enable a 100% supply of the global demand on electric energy and primary energy in 2032, and many years beyond (Appendix 1 in in (Kratzenberg, 2021)). It is shown that all continents present either (i) large deserts, where photovoltaic power plants can be installed; or (ii) remarkably high offshore potential for wind power plants, as is the case for Europe. Covering only 1.8%, e.g., of this desert space could generate the energy necessary for the expected global demand on electric energy demand in 2030. This estimation considers an average power factor of the globally installed PV systems of 26.1% (Appendix 1 in in (Kratzenberg, 2021)). This space reduces to 6.38 % to generate the global primary energy, e.g., solely with photovoltaic plants in 2030, which reduces by 42.5%, because of higher efficiencies in energy-to-work conversion (23%), the elimination of mining and refining activities related to coal, oil, gas and uranium exploration (12.7%), and policy-driven energy efficiency measures (6.9%) (Jacobson et al., 2017).

In the combination with wind and water this surface decreases furthermore. Alternatively, PV modules can also be integrated or applied to rooftops of residential and industrial buildings, which also decrease the necessary desert space. However, the cost of rooftop installations is ~ US$ 65/MWh slightly higher if compared to utility-scale systems ~ US$ 35/MWh, as estimated in (Klassen, 2020) for Australia. But rooftop installations result in a reduced actualization of the power grid, as its connection power is relatively low, if compared to further consumers. Therefore, residential installations in a large scale allow a very rapid expansion of the generation ability of a renewable power supported power grid. Such a rapid expansion ability is of strategic value in an exponential growth scenario to gain time, up to the point when the necessary construction works for power lines to access deserts and areas with a high wind energy potential are accomplished.

## 2.4 Transmission lines

We consider that the construction of necessary transmission lines does not have to delay the exponential expansion of the cumulative power of VRE power plants, if the power grid expansion is planed adequately in different countries, and by appropriate methods that confer to the predicted capacity of this exponential growth.

Most favorable locations with high wind or photovoltaic generation potentials are often localized in areas where no transmission lines are available, and where the demand on electric energy is also low (Georgilakis, 2008). Therefore, the availability of the necessary transmission lines for wind park installations can represent a barrier for the wind power development (Georgilakis, 2008). While PV power plant installations are not dependent on sites with high solar irradiation indices, cost optimization of very large plant in the can result in the future to the situation where the most of the PV power plants are installed in locations with high solar irradiation incidences. Therefore, at rigor in a minor degree also to PV power plant development could be restrained by not available transmission lines.

In Brazil a traditional workflow of the expansion of the generation capacity of the power grid, considers first the bidding of, e.g., a hydropower plant, after which the bidding of the necessary transmission line follows. Especially in the planning of the transmission lines for hydropower plants this is important because the company, which gains the bidding of the hydropower plant can evaluate first its source availability, which in most cases appears at one single local, or along a considered river. Once the hydropower plant bidding is accomplished the bidding of the necessary transmission line follows, which connect the hydropower plant to the national power grid. Such a concept can lead however to large time lags in the implementation of an exponential growth of VRE photovoltaic and wind power plants, because of the spatial distribution of its generators.

These time lags appear, as many different VRE projects, as accomplished by different companies use the same transmission line, because the generated energy is collected from a larger area, as in the case of the hydropower plant. Therefore, the many auctions related lead to time lags and the concept of a mores swift transmission planning is necessary, which can act even without the related auctions.

In the installation of hydropower plants or further conventional power plants the time lags in the transmission lines implementation, also present because of the necessary land appropriations and allowances necessary for the installation of transmission lines, were not perceived because of its pipeline workflow implementation. E.g., the pipeline construction of a hydropower dam in various concrete layers result in several years of construction time for large hydropower plants.

This pipeline construction concept is not present in VRE power plants, which can be installed in a parallel rather than a pipeline workflow to the advantage of rapid implementation. The parallel workflow allows a swift parallel installation of VRE power plants result in shorter time for its commission if the necessary power grid would be available. As a result, in a presently not tuned general work flow such VRE plants can stay many months without connection to the power grid, which creates losses. Therefore, a rapid expansion with lower cost should consider an improved workflow to provide the necessary infrastructure of power lines for the VRE plants. One proposal to avoid such time lags is a change in the bidding sequence. This means, before the bidding of the solar and wind power plants, a bidding of the power lines to a very resource rich land area is launched, already years before the installation of VRE plants are considered. However, the necessary investment should be adequately organized in such a case.

Effectively, in an exponential growth scenario an adequate planning of the transmission lines is necessary, which results in connections with adequate transmission capacities between windy regions and centers with high populations, or industrial centers where the most of the electric energy is demanded. Meanwhile, in most deregulated power markets, existing transmission planning practices do not look ahead towards expanding transmission grid to serve high wind resources. This restrain happens as the individual wind projects, at the beginning of the exploitation of the considered area with high wind potential cannot afford to pay for a major expansion of the power grid ((Georgilakis, 2008), citing (Piwko et al., 2005)). Therefore, wind resources remain underdeveloped and unexploited.

Meanwhile, the so called economic transmission planning (Piwko et al., 2005) represents the opportunity to access remote wind resources on the large scale of approximately 10 GW installation power, by the assembling of several wind projects in an expansion plan. However, there are some challenges related to this planning concept, such as its combination with state-of-the-art transmission planning for reliability of the power grid. Furthermore, market participants are still not required to contribute financially in economic transmission planning, as is the case for reliability planning. Additionally the economic transmission planning studies are more cost intensive as reliability planning. While the challenges are significant they are not insurmountable as discussed in (Georgilakis, 2008).

## 2.5 Skilled labor constraints

Considering an adequate anticipated planning of the training of skilled work force, we do not see that failing labor force can result in a constraint that limits the exponential growth of VRE technologies. We note however, that the training has a time lag as needed to prepare work force. Therefore, it must be oriented on an the predicted exponential growth, because experiences tell that growth in VRE systems can be limited by failing skilled labor force (Jagger et al., 2013). Governments, in combination with educational institutions play a major role in avoiding expansion shortages as a function of insufficient skilled labor force for installation and maintenance (Moorthy et al., 2019). Especially, in an exponential growth scenario VRE plants can installed at lower costs, if skilled labor forces are locally available, which results in employment benefits in many countries. Apart from the lower costs of its generated energy, renewable energy create significant more jobs per generated MWh, as in comparison to fossil fuels (Gonzalez, 2017). As evaluated by Jacob the transformation to an 100 % renewable energy generation lead worldwide to net employment benefit of 24.26 million jobs, as calculated for different countries (Jacobson et al., 2017). In this scenario, 25.39 million new fixed full time jobs are created in the construction and manufacturing industry, and 26.61 million jobs are created in the maintenance of the installed power plants, while 27.74 million jobs are lost in the nuclear and the fossil fuel industry.

## 2.6 Storage constraints

A future power system with an extremely high fraction of VRE power plants has to rely on large energy storages, because of the variability of the power flow of VRE generators. Otherwise, this variability leads to additional costs, because of the higher demand on the so called (i) peak power and (ii) super peak power plants (Hirth et al., 2015). However, we consider that the cost of the necessary storage for a 100% renewable energy grid with high VRE fractions is not a limiting factor for the exponential growth of solar and wind power plants.

Ultra-low-cost storage thermal storage units, as used for the room acclimatization and hot water production of buildings, present costs down to only US$ 1 / kWh (Jacobson et al., 2015). Such units are able to simulate or emulate electrical storage units, by its remote switching via demand response (DR) control for the

compensation of the variability of VRE power plants (Jacobson and Delucchi, 2011). Furthermore, the variability of the generated power of VRE power plants can be reduced because of the spatial and source type complementary generation of these plants.

Additionally, VRE day ahead, or intra-day forecasts (Perez et al., 2014), (Nobre, 2015), (Heinemann et al., 2006), which activate dispatchable generators and consumers, e.g., hydropower plants, and the DR controlled systems, also result in a lower demand for storage capacities.

Countries with a high fraction of hydropower plants present a large inherent storage of its electric power system, usable to compensate the variability of the VRE generation. In periods of high VRE generation the hydropower dispatch can be reduced, which increases the stored energy in the hydropower plant's water storage. This energy is then available in dry seasons or years at which the hydropower dispatch is increased.

The so named off-river hydropower plants are a reliable solution in countries with low hydroelectric power generations, and limitations in its power grid of transmission lines, in order to provide large dispatchable storages at several different locations for its power grid. While lithium-ion batteries present costs in the range of US$ 100 to 2500 per kWh the cost of pumped hydropower plants is with US$ 5 to 100 per kWh significantly lower in comparison (Koohi-Fayegh and Rosen, 2020). The availability of off-river pumped hydropower plants is also not a problem. As evaluated by shapes of the earth's surface, using a geographic information system (GIS), the worldwide availability of appropriate valleys for off-river-hydropower plants is two order of magnitude higher than the global demand in the case of an 100% renewable electrical power grid (Blakers et al., 2019). Such a large storage potential is more than sufficient to provide the necessary storage capacity to provide the primary energy in 2030 by VRE power plants, as the estimated primary energy demand in 2030 is solely 3.05 times higher than the estimated demand on electric energy, considering an improved efficiency in energy use (Appendix 1 in Kratzenberg 2021). Evaporation effects in especially dry regions can be reduced by floating photovoltaic power plants. Even in not so dry regions, such a concept can lead to a ~ 20% increase of the capacity factor of hydropower plants if only ~ 10% of its surface is covered with photovoltaic plants, as shown for in a simulation for the European hydropower plants (Quaranta et al.). Such a reduction is based on the effect that the floating PV plants reduce the lake's natural evaporation by 70% at the covered area {Quaranta, 2021 #473}.

## 2.7 Grid stability and integration

We consider the grid stability as not being a limiting factor in a 100% renewable power grid relied on high fractions of VRE energy generation. One of the challenges of a fully renewable power grid, is its stable operation of the power grid, because of the failing control functions of the wind and photovoltaic power plants in terms of the electric parameters of the power grid. Especially voltage, frequency and the current harmonics are important, control parameter. As based on synchronous generators the variation of these parameters are maintainable in specified limits. However, in the case of high renewable fractions a 100% renewable generation is mainly based on the power flow grid injection by the inverters of the VRE power plants. Such an injection presents, however, also an advantage which cannot be delivered by synchronous inverter. E.g., the inverters are able to measure the grid harmonics and as a result of its smart control inject an exactly tuned harmonic current that compensates these grid harmonics. As a result a highly stable power gird is configured, as based on smart VRE injection (Todeschini, 2010). The delivery of the necessary short circuit current in order to actuate the power grid's protective devices is also a problem as the used synchronous generators have a short circuit current, which is typically many times higher than its nominal current. However, the power grid of countries which present high fractions of renewable energy generations, such as Germany e.g., present an order of magnitude lower grid outage time as the power grids in countries with low fractions of such a generation (Gonzalez, 2017), but this is not a rigorous proof that VRE systems lead to higher stable power grids. Meanwhile, by further concepts the use of VRE systems can improve a power grid's stability in using technologies such as (i) Volt/volt ampere reactive (VAR) regulation; (ii) VRE converters with ride-trough capabilities; (iii) careful designed VRE controllers, which provide faster primary, secondary and tertiary response. The advantage of these systems is that they even provide a faster response speed than synchronous generators (Kroposki et al., 2017), which leads to an improved grid stability in principle.

Additionally, high short-circuit currents can be emulated by (i) virtual synchronous machines (VSM), and (ii) the use of synchronous machines, converted to providers of short circuit current. The latter are, e.g., deactivated machines from thermal power plants, which are operated as a motor, and provide in the case of a short circuit, a high current injection into the power grid. The VSM can inject power as stored in a battery in a short time frame to provide the necessary short circuit current. Furthermore, the use of intelligent protective devices of transmission lines can also result in a low cost solution, as they have a lower demand on the short circuit current for the trapping of the power grid's protective devices (Partain and Fraas, 2015).

Future power grids with extremely high VRE fractions can have short term grid support, as provided by stationary and electric car batteries, when connected to its charging stations. As discussed in (Partain and Fraas, 2015) cars are on average 70% of the time located in residential our city parking lots. At these locations, its energy storages can be recharged, considering that a charging infrastructure is provided. The temporary reduction of the car's charging current, or even an injection to the power grid in very short time frame of some seconds or minutes can potentially improve the grid's stability too.

**2.8 Investments**

We consider investments as not a barrier that can avoid the exponential growth of renewable power plants. The financial markets are presently strongly overvalued, as can be seen, e.g., by the Buffet indicator, which leads to large investment risks and uncertainties of potential losses in the investment market. Therefore, principally institutional investor organizations, which hold, e.g., social security assets, search safe investments with long term continuous return, which are characteristic for the investment in VRE plants. Such institutions are able to bankroll the expansion at a scale much larger than its initial medium scale expansion that only was possible because of the subsidies by governments (Sivaram, 2018). Furthermore, the valuation of VRE investments will increase significantly in the future because of the ability to generate energy at exceptionally low costs, and because of the exponential growth of these power plants. As a result of the steady cost reduction higher profits are obtained, which also gain the attention of large private investors. Meanwhile, especially third-country investments are still conceived as higher risk investments that dispel investors. However, special investment tools can reduce these risk too. In a co-investment with institutions that already have large experiences in VRE investments, such as the World-Bank, or the Inter-American Development bank, private or institutional investors can reduce its investment risk because of adequate due diligence procedures and execution. Additionally, the risk as related to the country's currency variation can be avoided by hedging this variation for investors in VRE power plants, as provided, e.g., by a country's political leadership (Sivaram, 2018). As a result large investments to enable exponential growth in development countries are also enabled. Internationally adopted carbon taxes, with not too low values per generated kg of $CO_2$ is realistic because of the related risks (Appendix 2 in in (Kratzenberg, 2021)), and will help to attract investment capital which will accelerate mass production and price decline per generated MWh.

**2.9 Energy policies**

In Brazil the energy market in divided principally in two different segments the regulated and the free market segment (Campos et al., 2020). Power source selective energy auctions in a regulated energy marked can by politic decision restrain the VRE fraction of utility scale wind and PV power plants operated in a power grid. Alternatives where such a limitation is not happen is the free energy market segment, or for residential PV installations, but these items contribute typically to a lower fraction of the energy generation.

Because VRE power plants show clearly the advantages of lower cost, higher employment, and climate change mitigation, we see no constraints because of the countries' energy policies, once the storage (Section 2.6) and grid stability (Section 2.7) constraints are attended. Furthermore, we think that it is of the interest of the global community, not only to mitigate climate change, and its perilous predicted consequences (Appendix 2 in (Kratzenberg, 2021)), but also generate its necessary energy at the lowest possible cost, provide employment and terminate city pollutions. Because the generation cost of wind and photovoltaic power plants decreases exponentially as a function of the global shipping of these plants (Figure 8 in (Kratzenberg, 2021)), the more these technologies are used globally the lower is the related generation cost, which should motivate the international cooperation and knowledge transfer in-between countries to benefit from the cost advantage of exponential or even higher than exponential growth of the use these technologies. Furthermore, there are no negative side effect to the large scale use of VRE technologies as in comparison to nuclear power plants in the case of nuclear accidents, or power plants operated with fossil fuel, because of its $CO_2$ emissions. Furthermore, VRE power plants provide what is of the most concern of voters, which is employment. Therefore policies should incentivize the use of these power plants providing sufficient energy auctions that enable the participation of these power plants.

## 2.10 Bidding price

In free energy markets the price at which an VRE plant generate energy is the so called bidding tariff, which shows an exponential decrease over time as observed for PV power plants in different countries (Arndt et al., 2019). The opening price of an energy auction, also called as the initial bidding price, is the lowest price of an organized energy bidding by a country's provider of electrical energy. If this stipulated opening price of a VRE auction is too low, without considering a discount (Porrua et al., 2010), then an auction can terminate frustrated, saying no company is available to accept to generate at this low energy price, and business is not happening. Another situation is when the energy provider specifies solely a specific power plant which is, however, not renewable. E.g., energy provider might be overcautious in relation to problems as related to the variability of power flow of VRE plants, which can result in such non-renewable auctions. In such a case the business with the VRE plants do also not happen, even if they can generate electric energy at much lower costs. Therefore, governments supported by experts, should determinate in advance the maximal VRE fractions in a power grid as a function of its physical and emulated storage capacities. In temporarily fashion this can also be based on the basis of the available peaker and super-peaker thermal power plants, but it has remembered that hydropower plants can also act as peaker power plants because of short time ramp for its power regulation.

In a bidding of the energy generation, the government's institution that buys the generated electoral energy makes usually a contract over a long time period of up to 20 years, e.g., for the bought energy. Long term contracts based on fossil fuel generation do not enable a 100% renewable energy generation in the here predicted time frame. Furthermore, because of the strong price decrease its energy will cost much more than the energy generated by VRE plants. However, the companies who operate natural gas supplied power plants have usually two earnings. The first earning they obtain without the generation of energy only by bringing available a possible generation, and the second earning they obtain for the generated energy as contracted. In an exponential VRE expansion such power plants are important to compensate for the variability in order to allow an exponential increase of VRE plants until the necessary storage capacity is installed. Furthermore, advanced business models should be considered too for the thermal generators, as even without the generation of energy the inertia of these power plants' synchronous generators can provide the presently still necessary large short-circuit current in a power grid, which activates its protective devices in a case of failure as discussed in Section 3.7.

## 3 Conclusions

Because of the historical exponential cost reduction of the energy generated by wind and photovoltaic plants power plans, these plants generate energy at competitive costs presently on average. Therefore at most of the installation locations VRE power plants generate energy at similar or even lower costs in comparison to conventional power plants. Therefore, and because of the expected further cost reductions, subsidies, as were necessary in the past for these VRE power plants, are not anymore necessary in the future for most of the installation locations. However, different conditions can halt back the exponential growth of VRE power plant in the future, which we discussed here and we think that probably the transmission line availability is probable the strongest issue that can restrain the installation of wind and in the future also large scale photovoltaic power plants. We think that it is important to develop activities that enable, rather than restrain, the exponential growth of VRE power plants because of the many related benefits. The actualization of the transmission line capacity for the admission of large scale capacities of VRE plants it also important in the case when the primary energy is to the most part in generated by VRE power plants as representing a 100% supply by renewable energy in 2030. Such problems appear after 2025 or before, when the power grids are supplied by 100% of renewable energy.

An appropriate actualization of the power grid should not only be accomplishable in shortest time frame but it should also be based on low cost energy storages such as thermal storages controlled by DR, or off-river pumped hydro power plants, avoiding the increase the consumers' electric energy price. Too high storage costs as related to lithium ion batteries can by its own restrain an exponential grow of VRE power plants. Another important part are energy policies in different countries, which should provide the necessary bureaucratic and legislative background that allows an exponential growth, rather than hold back such a growth. Such a background is important considering the benefits of the VRE expansion and the severity of climate change in an otherwise linear expansion scenario, which contribution is of minor importance. Specifically such an enabled growth results in (i) reduced energy generation costs, not immediately but after the necessary storage and infrastructure is payed, (ii) increased employment, (iii) cities with clean air and less respiratory problems, and finally (iv) the reduction of $CO_2$ in the atmosphere, which eventually can still avoid the most severe outcomes of future climate change.